\documentclass[12pt,a4paper,onecolumn]{IEEEtran}
\usepackage{geometry}
\usepackage{amsmath,amssymb,amsthm,mathrsfs,amssymb,mathrsfs,txfonts}
\usepackage{latexsym,color,booktabs}
\usepackage[mathscr]{eucal}
\theoremstyle{definition}

\begin{document}
\begin{center}
\LARGE\bf{Ordering states with various coherence measures}
\end{center}

\centerline{Long-Mei Yang$^1$, Bin Chen$^2$, Shao-Ming Fei $^1$, Zhi-Xi Wang$^1$}

\centerline{$^1$School of Mathematical Sciences, Capital Normal University, Beijing 100048, China}
\centerline{$^2$School of Mathematical Sciences, Tianjin Normal University, Tianjin 300387, China}

\bigskip

\noindent\textbf{Abstract}

Quantum coherence is one of the most significant theories in quantum physics.
Ordering states with various coherence measures is an intriguing task in quantification theory of coherence.
In this paper, we study this problem by use of four important coherence measures -- the $l_1$ norm of coherence, the relative entropy of coherence, the geometric measure of coherence and the modified trace distance measure of coherence.
We show that each pair of these measures give a different ordering of qudit states when $d\geq 3$.
However, for single-qubit states, the $l_1$ norm of coherence and the geometric coherence provide the same ordering.
We also show that the relative entropy of coherence and the geometric coherence give a different ordering for single-qubit states.
Then we partially answer the open question proposed in [Quantum Inf. Process. 15, 4189 (2016)] whether all the coherence measures give a different ordering of states.

\noindent\textbf{Keywords} $l_1$-norm of coherence, relative entropy of coherence, geometric measure of coherence, modified trace distance of coherence, ordering states.

\noindent\textbf{Introduction}

Quantum coherence is one of the most outstanding features in quantum mechanics.
It is very essential in various research fields such as low-temperature thermodynamics \cite{1,2,3,4,5}, quantum biology \cite{6,7,8,9,10,11}, nanoscale physics \cite{12,13}, etc.
Although formulating resource theory of quantum coherence is a long-standing open problem, it has only been proposed by Baumgratz {\it et al.} recently \cite{Baumgratz}.
In their seminal work, conditions that a suitable measure of coherence should satisfy have been put forward.
After that, many efforts have been made in quantification of coherence.
Up to now, various proper quantifiers have been given, such as the $l_1$ norm of coherence, the relative entropy of coherence \cite{Baumgratz}, the geometric measure of coherence \cite{Streltsov} and the modified trace distance measure of coherence \cite{Tong,Chen}, etc.

Based on various physical contexts, different values of coherence may reflect different properties of quantum states.
Generally, one cannot say that the coherence of a state $\rho_1$ is smaller than that of $\rho_2$, since different coherence measures may provide a different ordering for these two states.
Similar to the case of quantum entanglement, different measures of quantum coherence characterize different aspects of the state,
and play different roles in quantum information processing. Hence, a given quantum state may behavior better in one information processing, but worse in another information task.
On the other hand, one can identify two measures of coherence to some extent if they give the same ordering for all quantum states.
Therefore, it is worthy of a study on the ordering of quantum states under different measures of quantum coherence.
It should be noted that this issue has been extensively investigated in the theory of quantum entanglement.
In \cite{entanglement}, Virmani {\it et al.} showed that any two entanglement measures placing the same ordering on states must be identical, as long as they coincide on pure states.
However, less has been known for quantum coherence.
In this paper, we focus on ordering states based on various coherence measures.

In \cite{Liu}, Liu {\it et al.} consider the $l_1$ norm of coherence and the relative entropy of coherence, and show that these two measures do not give the same ordering of states.
Then they propose an open question: whether all the coherence measures give a different ordering of states?
That is to say, whether there exist quantum states $\rho_1$ and $\rho_2$, such that the following relation fails for any two coherence measures $\mathcal{C}_A$ and $\mathcal{C}_B$: $\mathcal{C}_A(\rho_1)\leq\mathcal{C}_A(\rho_2)$ iff $\mathcal{C}_B(\rho_1)\leq\mathcal{C}_B(\rho_2)$.
In this paper, we investigate this problem.
We mainly focus on four coherence measures -- the $l_1$ norm of coherence, the relative entropy of coherence, the geometric measure of coherence and the modified trace distance measure of coherence.
We show that each pair of these measures do not give the same ordering of high-dimensional states in general.
However, the $l_1$ norm of coherence and the geometric coherence provide the same ordering for single-qubit states,
while the relative entropy of coherence and the geometric coherence still give rise to a different ordering in this case.
Thus we partially answer the open question proposed in \cite{Liu}.
Additionally, we provide some special sets of quantum states such that each pairs of the four coherence measures give the same ordering.

This paper is organized as follows.
In Sect. 2, we review some basic concepts about quantification theory of coherence and some coherence measures we will use in this paper.
In Sect. 3, we present our main results via detailed examples by using pairwise coherence measures.
We also extend our discussion to the coherence of formation and the Tsallis relative $\alpha$-entropy of coherence for single-qubit states.
We conclude our results in Sect. 4.

\medskip

\noindent\textbf{The quantification of coherence}\smallskip

In this section, we first review some basic concepts about quantification of coherence.

For a given $d$-dimensional Hilbert space $\mathcal{H}$, let us fix an orthonormal basis $\{|i\rangle\}_{i=1}^d$.
Then the incoherent states are defined as:
\begin{equation}
\sigma=\sum\limits_{i=1}^d p_i|i\rangle\langle i|,
\end{equation}
where $p_i\geq0, \ \Sigma_{i=1}^dp_i=1$.
The set of all the incoherent states is denoted as $\mathcal{I}$.
Let $\Lambda$ be a completely positive trace preserving (CPTP) map:
\begin{equation}
\Lambda (\rho)=\sum\limits_{n}K_n\rho K_n^{\dag},
\end{equation}
where $\{K_n\}$ is a set of Kraus operators satisfying $\Sigma_nK_n^{\dag}K_n=\mathbb{I}_d$, with $\mathbb{I}_d$ the identity operator.
If $K_n\mathcal{I}K_n^{\dag}\subseteq \mathcal{I}$ for all $n$, then we call $\{K_n\}$ a set of incoherent Kraus operators,
and the corresponding $\Lambda$ an incoherent operation.

In Ref. \cite{Baumgratz}, Baumgratz {\it et al.} proposed a resource-theoretic framework for quantifying quantum coherence.
Any function $\mathcal{C}$ defined on a space of quantum states can be employed as a proper measure of coherence,
if it satisfies the following four conditions :

{\rm (B1)} $\mathcal{C}(\rho)\geq0$, $\mathcal{C}(\rho)=0$ if and only if $\rho\in\mathcal{I}$;

{\rm (B2)} $\mathcal{C}(\Lambda(\rho))\leq\mathcal{C}(\rho)$ for any incoherent operation $\Lambda$;

{\rm (B3)} $\Sigma_np_n\mathcal{C}(\rho_n)\leq \mathcal{C}(\rho)$, where $p_n={\rm Tr}(K_n\rho K_n^{\dag}), \ \rho_n=K_n\rho K_n^{\dag}/p_n$,
$\{K_n\}$ is a set of incoherent Kraus operators;

{\rm (B4)} $\mathcal{C}(\Sigma _i p_i\rho_i)\leq\Sigma _i p_i\mathcal{C}(\rho_i)$ for any
set of quantum states $\{\rho_i\}$ and any probability distribution $\{p_i\}$.

Recently, Yu {\it et al.} put forward an alternative framework for quantifying coherence \cite{Tong}.
This framework is equivalent to the previous one proposed by Baumgratz \emph{et al.}\cite{Baumgratz}.
A nonnegative function $\mathcal{C}$ can be used as a measure of coherence, if it satisfies:

{\rm (C1)} $\mathcal{C}(\rho)\geq0$, $\mathcal{C}(\rho)\geq0$ if and only if $\rho\in\mathcal{I}$;

{\rm (C2)} $\mathcal{C}(\Lambda(\rho))\leq\mathcal{C}(\rho)$ for any incoherent operation $\Lambda$;

{\rm (C3)} $\mathcal{C}(p_1\rho_1\oplus p_2\rho_2)=p_1\mathcal{C}(\rho_1)+p_2\mathcal{C}(\rho_2)$ for block diagonal states
$\rho$ in the incoherent basis.

In accordance with the above frameworks, several legitimate coherence measures have been provided so far.
In this paper, we mainly consider four coherence measures -- the $l_1$ norm of coherence, the relative entropy of coherence, the geometric measure of coherence and the modified trace distance measure of coherence.

The $l_1$ norm of coherence is defined as
\begin{equation}\label{3}
\mathcal{C}_{l_1}(\rho)=\sum\limits_{i\neq j}|\rho_{ij}|,
\end{equation}
where $\rho_{ij}=\langle i|\rho|j\rangle$.

The relative entropy of coherence is defined as
\begin{equation}\label{rel}
\mathcal{C}_r(\rho)=\underset{\sigma\in\mathcal{I}}{\min}\mathcal{S}(\rho\|\sigma)=\mathcal{S}(\rho_{diag})-\mathcal{S}(\rho),
\end{equation}
where $\mathcal{S}(\rho\|\sigma)={\rm Tr}(\rho\log\rho-\rho\log\sigma)$ is the quantum relative entropy,
$\mathcal{S}(\rho)=- {\rm Tr}(\rho\log\rho)$ is the von Neumann entropy, and $\rho_{diag}=\Sigma_i\rho_{ii}|i\rangle\langle i|$.

The geometric measure of coherence is defined as
\begin{equation}
\mathcal{C}_g(\rho)=1-\underset{\sigma\in\mathcal{I}}{\max}F(\rho,\sigma),
\end{equation}
where $F(\rho,\sigma)=\Big({\rm Tr}\sqrt{\sqrt{\sigma}\rho\sqrt{\sigma}}\Big)^2$ is the fidelity of two density operators $\rho$
and $\sigma$.
When $\rho$ is a pure state, $C_{g}(\rho)=1-\max_{i}\{\rho_{ii}\}$, where $\rho_{ii}=\langle i|\rho|i\rangle$ \cite{Zhang}.

The modified trace distance measure of coherence is defined as
\begin{equation}
\mathcal{C}_{tr}^{\prime}(\rho)=\underset{\lambda\geq0,\delta\in\mathcal{I}}{\min}\parallel\rho-\lambda\delta\parallel_{tr}.
\end{equation}
It has been shown that $\mathcal{C}_{tr}^{\prime}(\rho)=\mathcal{C}_{l_1}(\rho)$ if $\rho$ is a single-qubit state \cite{Chen}.

It should be noted that the last two coherence measures have no analytical expressions in general.
However, for some special classes of coherent states, explicit formulae have been presented in \cite{Zhang,Chen}.
For example, for the maximally coherent mixed states (MCMS) \cite{Singh},
\begin{equation}\label{mcms}
\rho_m=p|\phi_d\rangle\langle\phi_d|+\frac{1-p}{d}\mathbb{I}_d,
\end{equation}
where $0<p\leq1$, and $|\phi_d\rangle=\frac{1}{\sqrt{d}}\sum_{i=1}^{d}|i\rangle$ is the maximally coherent state,
one has $\mathcal{C}_g(\rho_m)=1-[\sqrt{1-p}+\frac{1}{d}(\sqrt{1-p+dp}-\sqrt{1-p})]^2$ \cite{Zhang},
and $\mathcal{C}_{tr}^{\prime}(\rho_m)=p$ \cite{Chen}.

\medskip

\noindent\textbf{Ordering states with coherence measures}\smallskip

Let us recall the concept of ordering states with coherence measures.
We first note that all the states can be ordered under a coherence measure $\mathcal{C}$, since $\mathcal{C}(\rho)$ is always a nonnegative real number.
Then a natural question is raised: are there any two coherence measures $\mathcal{C}_A$ and $\mathcal{C}_B$ which give rise to the same ordering of all states?
Here the same ordering means that the following relation holds for any two states $\rho_1$ and $\rho_2$:
\begin{equation}
\mathcal{C}_A(\rho_1)\leq\mathcal{C}_A(\rho_2)\Leftrightarrow\mathcal{C}_B(\rho_1)\leq\mathcal{C}_B(\rho_2).
\end{equation}
Otherwise, we say that the two measures give a different ordering.

In this section, we discuss ordering states with pairs of coherence measures among $\mathcal{C}_{l_1}$, $\mathcal{C}_r$, $\mathcal{C}_g$ and $\mathcal{C}_{tr}^{\prime}$ via detailed examples.
We will show that $\mathcal{C}_{l_1}$, $\mathcal{C}_r$, $\mathcal{C}_g$ and $\mathcal{C}_{tr}^{\prime}$ generate a different ordering of qudit $(d\geq3)$ states.
For single-qubit states, we show that $\mathcal{C}_{l_1}$ and $\mathcal{C}_g$ give the same ordering,
while $\mathcal{C}_r$ and $\mathcal{C}_g$ provide a different ordering.

\medskip

\noindent\textbf{Ordering states with $\mathcal{C}_{l_1}$ and $\mathcal{C}_g$}\smallskip

We first consider two-dimensional quantum systems.
Any density operator acting on a two-dimensional quantum system can be generally written as
\begin{equation}
\rho=\left(
\begin{array}{cc}
  a & b \\
  b^\ast & 1-a
\end{array}
\right),
\end{equation}
where $|a|^2+|b|^2\leq1$.
Then we have $\mathcal{C}_{l_1}(\rho)=2|b|$ and $\mathcal{C}_g(\rho)=\frac{1-\sqrt{1-4|b|^2}}{2}$ \cite{Zhang}.
It can be seen that $\mathcal{C}_{l_1}(\rho)$ and $\mathcal{C}_{g}(\rho)$ are both increasing functions with respect to $|b|$.
Thus, for all single-qubit states, the coherence measures $\mathcal{C}_{l_1}$ and $\mathcal{C}_{g}$ give the same ordering,
since $\mathcal{C}_{l_1}(\rho_1)\leq\mathcal{C}_{l_1}(\rho_2)\Leftrightarrow|b_1|\leq|b_2|\Leftrightarrow
\mathcal{C}_{g}(\rho_1)\leq\mathcal{C}_{g}(\rho_2)$,
where $\rho_1=\left(
\begin{array}{cc}
  a_1 & b_1 \\
  b_1^\ast & 1-a_1
\end{array}
\right)$
and
$\rho_2=\left(
\begin{array}{cc}
  a_2 & b_2 \\
  b^\ast_2 & 1-a_2
\end{array}
\right)$
are arbitrary single-qubit states.

We now discuss the case of high-dimensional quantum systems.
Let $|\psi\rangle=\sum_{i=1}^{d}\sqrt{\lambda_{i}}|i\rangle$ and $|\phi\rangle=\sum_{i=1}^{d}\sqrt{\mu_{i}}|i\rangle$ be two pure states,
where $\lambda_{i}\geq0,~\sum_{i=1}^{d}\lambda_{i}=1$, and $\mu_{i}\geq0,~\sum_{i=1}^{d}\mu_{i}=1$.
Then we have $\mathcal{C}_{l_1}(|\psi\rangle)\leq\mathcal{C}_{l_1}(|\phi\rangle)\Leftrightarrow\sum_{i=1}^{d}\sqrt{\lambda_{i}}\leq\sum_{i=1}^{d}\sqrt{\mu_{i}}$,
and $\mathcal{C}_{g}(|\psi\rangle)\leq\mathcal{C}_{g}(|\phi\rangle)\Leftrightarrow\max_i\{\lambda_i\}\geq\max_i\{\mu_i\}$.
Thus $\mathcal{C}_{l_1}(|\psi\rangle)\leq\mathcal{C}_{l_1}(|\phi\rangle)\Leftrightarrow\mathcal
{C}_{g}(|\psi\rangle)\leq\mathcal{C}_{g}(|\phi\rangle)$
if the two conditions $\sum_{i=1}^{d}\sqrt{\lambda_{i}}\leq\sum_{i=1}^{d}\sqrt{\mu_{i}}$ and $\max_i\{\lambda_i\}\geq\max_i\{\mu_i\}$ hold at the same time.

Let us consider a special case where $\lambda_{1}=\lambda\geq0,~\lambda_{2}=\lambda_{3}=\cdots=\lambda_{d}$, i.e., $|\psi\rangle=\sqrt{\lambda}|1\rangle+\sqrt{\frac{1-\lambda}{d-1}}\sum_{i=2}^{d}|i\rangle$.
Then we have $\mathcal{C}_{l_1}(|\psi\rangle)=(\sqrt{\lambda}+\sqrt{(d-1)(1-\lambda)})^2-1$.
Note that $\mathcal{C}_{l_1}(|\psi\rangle)$ is an increasing function with respect to $\lambda$ when $\lambda\leq\frac{1}{d}$,
while a decreasing function when $\lambda\geq\frac{1}{d}$.
Let $|\phi\rangle=\sqrt{\mu}|1\rangle+\sqrt{\frac{1-\mu}{d-1}}\sum_{i=2}^{d}|i\rangle$.
We consider the following cases:

(i) If $\lambda\leq\frac{1}{d},~\mu\leq\frac{1}{d}$, then we have
$\mathcal{C}_{l_1}(|\psi\rangle)\leq\mathcal{C}_{l_1}(|\phi\rangle)\Leftrightarrow\lambda\leq\mu
\Leftrightarrow\mathcal{C}_{g}(|\psi\rangle)=1-\frac{1-\lambda}{d-1}\leq\mathcal{C}_{g}(|\phi\rangle)=1-\frac{1-\mu}{d-1}$.
Thus the coherence measures $\mathcal{C}_{l_1}$ and $\mathcal{C}_g$ generate the same ordering in this case.

(ii) If $\lambda\geq\frac{1}{d},~\mu\geq\frac{1}{d}$, then we have
$\mathcal{C}_{l_1}(|\psi\rangle)\leq\mathcal{C}_{l_1}(|\phi\rangle)\Leftrightarrow\lambda\geq\mu
\Leftrightarrow\mathcal{C}_{g}(|\psi\rangle)=1-\lambda\leq\mathcal{C}_{g}(|\phi\rangle)=1-\mu$.
Thus $\mathcal{C}_{l_1}$ and $\mathcal{C}_g$ also generate the same ordering in this case.

(iii) If $\lambda\geq\frac{1}{d},~\mu<\frac{1}{d}$, then we have
$\mathcal{C}_{g}(|\psi\rangle)\leq\mathcal{C}_{g}(|\phi\rangle)\Leftrightarrow(d-1)\lambda\geq1-\mu$.
To find different ordering pairs, one may choose $\lambda$ and $\mu$ that satisfy
$\mathcal{C}_{l_1}(|\psi\rangle)>\mathcal{C}_{l_1}(|\phi\rangle)\Leftrightarrow
\sqrt{\lambda}+\sqrt{(d-1)(1-\lambda)}>\sqrt{\mu}+\sqrt{(d-1)(1-\mu)}$.
This implies that $\sqrt{1-\lambda}+\sqrt{1-\mu}>\sqrt{(d-1)\lambda}+\sqrt{(d-1)\mu}\geq\sqrt{1-\mu}+\sqrt{(d-1)\mu}$.
Hence $\frac{1-\mu}{d-1}\leq\lambda<1-(d-1)\mu,~d\geq3$, and in this case $\mathcal{C}_{l_1}$ and $\mathcal{C}_g$ generate a different ordering.
Therefore we conclude that the coherence measures $\mathcal{C}_{l_1}$ and $\mathcal{C}_g$ do not give the same ordering in $d$-dimensional quantum systems when $d\geq3$.
They can only provide the same ordering for families of quantum states.

As another example, let us consider $\rho_m$ defined in (\ref{mcms}).
We have that $\mathcal{C}_{l_1}$ and $\mathcal{C}_g$ provide the same ordering for this class of states,
since $\mathcal{C}_{l_1}(\rho_m)=(d-1)p$ and $\mathcal{C}_g(\rho_m)=1-[\sqrt{1-p}+\frac{1}{d}(\sqrt{1-p+dp}-\sqrt{1-p})]^2$
are both increasing functions with respect to $p$,
thus $\mathcal{C}_{l_1}(\rho_m)\leq\mathcal{C}_{l_1}(\widetilde{\rho_m})\Leftrightarrow p\leq\widetilde{p}\Leftrightarrow
\mathcal{C}_{g}(\rho_m)\leq\mathcal{C}_{g}(\widetilde{\rho_m})$,
where $\widetilde{\rho_m}=\widetilde{p}|\phi_d\rangle\langle\phi_d|+\frac{1-\widetilde{p}}{d}\mathbb{I}_d$.

\medskip

\noindent\textbf{Ordering states with $\mathcal{C}_{r}$ and $\mathcal{C}_g$}\smallskip

In Ref. \cite{Liu}, the authors have shown that $\mathcal{C}_{r}$ and $\mathcal{C}_{l_1}$ give rise to a different ordering
of single-qubit states.
Taking into account the previous result that $\mathcal{C}_{l_1}$ and $\mathcal{C}_g$ provide the same ordering of single-qubit states,
we have that $\mathcal{C}_{r}$ and $\mathcal{C}_{g}$ must provide a different ordering in this case.
Just like the discussion in \cite{Tsa}, to find $\sigma_1$ and $\sigma_2$ that satisfy both
$\mathcal{C}_r(\sigma_1)>\mathcal{C}_r(\sigma_2)$ and $\mathcal{C}_g(\sigma_1)<\mathcal{C}_g(\sigma_2)$,
one can choose $t_1$ and $t_2$ ($t_1<t_2$) such that
$H\Big(\frac{1-\sqrt{1-t_1^2}}{2}\Big)>1-H\Big(\frac{1-t_2}{2}\Big)$,
and then find $z_1$ and $z_2$ ($0\leq z_1,z_2\leq1$)
by using $H\Big(\frac{1}{2}-\frac{z_1}{2}\Big)-H\Big(\frac{1}{2}-\frac{\sqrt{z_1^2+t_1^2}}{2}\Big)>
H\Big(\frac{1}{2}-\frac{z_2}{2}\Big)-H\Big(\frac{1}{2}-\frac{\sqrt{z_2^2+t_2^2}}{2}\Big)$,
where $\sigma_1=\frac{1}{2}\left(
                           \begin{array}{cc}
                             1+z_1 & t_1 \\
                             t_1 & 1-z_1 \\
                           \end{array}
                         \right),
                         \sigma_2=\frac{1}{2}\left(
                           \begin{array}{cc}
                             1+z_2 & t_2 \\
                             t_2 & 1-z_2 \\
                           \end{array}
                         \right)$,
and $H(x)=-x\log x-(1-x)\log(1-x)$.
For instance, assume
\begin{equation}
\sigma_1=\left(
\begin{array}{cc}
  \frac{4}{5} & \frac{2}{5} \\
 \frac{2}{5} & \frac{1}{5}
\end{array}
\right), \ \ \sigma_2=\left(
\begin{array}{ccc}
  \frac{1}{2} & \frac{1}{\sqrt{6}}\\
   \frac{1}{\sqrt{6}} & \frac{1}{2}
\end{array}
\right).
\end{equation}
we have that $\mathcal{C}_{r}$ and $\mathcal{C}_{g}$ must provide a different ordering.
This can be seen from the fact that $\mathcal{C}_r(\sigma_1)=0.7219>\mathcal{C}_r(\sigma_2)=0.5576$, and
$\mathcal{C}_g(\sigma_1)=\frac{1}{5}<\mathcal{C}_g(\sigma_2)=\frac{3-\sqrt{3}}{6}$.

For high-dimensional quantum systems, let us define two $d$-dimensional states ($d\geq3$) as follows:
\begin{equation}
\sigma_1^{(d)}=p\sigma_1\oplus (1-p)\delta_1^{(d-2)},\ \sigma_2^{(d)}=p\sigma_2\oplus (1-p)\delta_2^{(d-2)},
\end{equation}
where $0<p\leq 1$, and $\delta_1^{(d-2)}$, $\delta_2^{(d-2)}$ are $(d-2)$-dimensional incoherent states.
Then $\mathcal{C}_r(\sigma_1^{(d)})=p\mathcal{C}_r(\sigma_1)>p\mathcal{C}_r(\sigma_2)=\mathcal{C}_r(\sigma_2^{(d)})$,
and $\mathcal{C}_r(\sigma_1^{(d)})=p\mathcal{C}_g(\sigma_1)<p\mathcal{C}_g(\sigma_2)=\mathcal{C}_g(\sigma_2^{(d)})$.
Thus, the coherence measures $\mathcal{C}_{r}$ and $\mathcal{C}_{g}$ give rise to a different ordering of arbitrary dimensional states.

However, for some special classes of states, $\mathcal{C}_{r}$ and $\mathcal{C}_{g}$ could generate the same ordering.
For instance, it has been shown that for all single-qubit states with a fixed mixedness,
the coherence measures $\mathcal{C}_{l_1}$ and $\mathcal{C}_{r}$ have the same ordering \cite{mix},
thus $\mathcal{C}_{r}$ and $\mathcal{C}_{g}$ also have the same ordering in this case, since
$\mathcal{C}_{l_1}$ and $\mathcal{C}_g$ provide the same ordering for all single-qubit states.
Let us consider again a class of MCMS $\rho_m$, one has
$\mathcal{C}_r(\rho_m)=\log d+\frac{1+(d-1)p}{d}\log \frac{1+(d-1)p}{d}+\frac{(d-1)(1-p)}{d}\log \frac{1-p}{d}$.
It can been seen that $\mathcal{C}_r(\rho_m)$ and $\mathcal{C}_g(\rho_m)$ are both increasing functions with respect to $p$,
hence give rise to the same ordering.

\medskip

\noindent\textbf{Ordering states with $\mathcal{C}_{tr}^{\prime}$ and $\mathcal{C}_g$}\smallskip

It is obvious that $\mathcal{C}_{tr}^{\prime}$ and $\mathcal{C}_{g}$ give the same ordering of single-qubit states, since in this case $\mathcal{C}_{tr}^{\prime}(\rho)=\mathcal{C}_{l_1}(\rho)$, $\mathcal{C}_{l_1}$ and $\mathcal{C}_{g}$ provide the same ordering.

For high-dimensional quantum systems,
since the two coherence measures $\mathcal{C}_{tr}^{\prime}$ and $\mathcal{C}_g$ have no analytical expressions in general,
we can only take into account special examples to show that they do not provide the same ordering.
To this end, let us consider two qutrit states $\rho_1=|\phi_2\rangle\langle\phi_2|\oplus 0$, and $\rho_2=p|\phi_3\rangle\langle\phi_3|+\frac{1-p}{3}\mathbb{I}_3$,
where $|\phi_2\rangle=\frac{1}{\sqrt{2}}\sum_{i=1}^2|i\rangle$, and $|\phi_3\rangle=\frac{1}{\sqrt{3}}\sum_{i=1}^3|i\rangle$.
Then we have $\mathcal{C}_{tr}^{\prime}(\rho_1)=\mathcal{C}_{tr}^{\prime}(|\phi_2\rangle\langle\phi_2|)=1, \ \mathcal{C}_{tr}^{\prime}(\rho_2)=p$,
$\mathcal{C}_{g}(\rho_1)=\mathcal{C}_{g}(|\phi_2\rangle\langle\phi_2|)=\frac{1}{2}$, $\mathcal{C}_{g}(\rho_2)=1-(\frac{2}{3}\sqrt{1-p}+\frac{1}{3}\sqrt{1+2p})^2$.
It can be seen that $\mathcal{C}_g(\rho_2)\leq\frac{2}{3}$, since
$\mathcal{C}_{g}(\rho_2)$ is an increasing function with respect to $p$.
Thus there exists a $p<1$ such that $\mathcal{C}_{g}(\rho_2)>\mathcal{C}_{g}(\rho_1)$.
For instance, let $\rho_2^{\prime}=\frac{99}{100}|\phi_3\rangle\langle\phi_3|+\frac{1}{300}\mathbb{I}_3$,
then we get $\mathcal{C}_{g}(\rho_2^{\prime})>\mathcal{C}_{g}(\rho_1)$, and
$\mathcal{C}_{tr}^{\prime}(\rho_2^{\prime})=\frac{99}{100}<\mathcal{C}_{tr}^{\prime}(\rho_1)=1$.
That is to say, $\mathcal{C}_{tr}^{\prime}$ and $\mathcal{C}_g$ provide a different ordering of qutrit states.
Using similar approach which transforms a $3\times3$ density matrix to a $d\times d$ density matrix ($d\geq3$) by direct sum of an incoherent state,
we have that the coherence measures $\mathcal{C}_{tr}^{\prime}$ and $\mathcal{C}_{g}$ generate a different ordering of qudit ($d\geq3$) states.

Similar to the above discussion, $\mathcal{C}_{tr}^{\prime}$ and $\mathcal{C}_g$ provide the same ordering for $\rho_m$,
since $\mathcal{C}_{tr}^{\prime}(\rho_m)$ and $\mathcal{C}_g(\rho_m)$ are both increasing functions with respect to $p$.

\medskip

\noindent\textbf{Ordering states with $\mathcal{C}_{tr}^{\prime}$ and $\mathcal{C}_{l_1}$}\smallskip

Before discussing ordering states with $\mathcal{C}_{tr}^{\prime}$ and $\mathcal{C}_{l_1}$, let us first provide an upper bound of $\mathcal{C}_{tr}^{\prime}$.
Let $\sigma$ be a pure qudit state and $\delta_i=|i\rangle\langle i|$, $1\leq i \leq d$.
Then we have
\begin{eqnarray*}
\mathcal{C}_{tr}^{\prime}(\sigma)
&\leq& \underset{\lambda\geq0,1\leq i\leq d}{\min}\parallel\sigma-\lambda\delta_i\parallel_{tr}\\
&=&\underset{\lambda\geq0}{\min}\sqrt{\lambda^2+(2-4\underset{i}{\max}\{\sigma_{ii}\})\lambda+1}\\
&=&
\left\{\begin{array}{ll} \sqrt{1-(2\underset{i}{\max}\{\sigma_{ii}\}-1)^2} & \text{if} \ \underset{i}{\max}\{\sigma_{ii}\}\geq\frac{1}{2} ,\\[3mm]
   1 & \text{if}\ \  \underset{i}{\max}\{\sigma_{ii}\}<\frac{1}{2}.\end{array}\right.
\end{eqnarray*}
Thus, for any qudit state $\rho$, $\mathcal{C}_{tr}^{\prime}(\rho)\leq\Sigma_ip_i\mathcal{C}_{tr}^{\prime}(\rho_i)\leq 1$,
where $\rho=\Sigma_ip_i\rho_i$ is any pure state decomposition of $\rho$ with $p_i\geq0, \ \Sigma_ip_i=1$.
For single-qubit states, since $\mathcal{C}_{l_1}(\rho)=\mathcal{C}_{tr}^{\prime}(\rho)$, $\mathcal{C}_{l_1}$ and $\mathcal{C}_{tr}^{\prime}$ of course provide the same ordering in this case.

Now consider the following pure qutrit states,
$|\psi\rangle=\sqrt{\lambda_1}|1\rangle+\sqrt{\lambda_2}|2\rangle$ and
$|\phi\rangle=\sqrt{\mu_1}|1\rangle+\sqrt{\mu_2}(|2\rangle+|3\rangle)$, where $\lambda_1+\lambda_2=1, \ \mu_1+2\mu_2=1$.
Assume that $\frac{1}{2}\leq\lambda_1<\mu_1<\frac{8}{9}$. Then we find
$\mathcal{C}_{tr}^{\prime}(|\phi\rangle)\leq\sqrt{1-(2\mu_1-1)^2}<2\sqrt{\lambda_1(1-\lambda_1)}=\mathcal{C}_{tr}^{\prime}(|\psi\rangle)$,
and $\mathcal{C}_{l_1}(|\phi\rangle)=(\sqrt{\mu_1}+\sqrt{2(1-\mu_1)})^2-1>1\geq\mathcal{C}_{l_1}(|\psi\rangle)$.
For instance, let $\rho_1$ and $\rho_2$ be two pure qutrit states,
$$\rho_1=\left(\begin{array}{ccc}
  \frac{1}{2} & \frac{1}{2} &0 \\
 \frac{1}{2} & \frac{1}{2}&0\\
 0 & 0&0
\end{array}
\right), \ \
\rho_2=\left(
\begin{array}{ccc}
  \frac{3}{4} & \frac{\sqrt{6}}{8} & \frac{\sqrt{6}}{8}\\
   \frac{\sqrt{6}}{8} & \frac{1}{8}& \frac{1}{8}\\
   \frac{\sqrt{6}}{8} & \frac{1}{8}& \frac{1}{8}
\end{array}
\right).$$
Using the previous results, one can easily get $\mathcal{C}_{tr}^{\prime}(\rho_2)\leq\sqrt{1-(2\cdot\frac{3}{4}-1)^2}=\frac{\sqrt{3}}{2}<\mathcal{C}_{tr}^{\prime}(\rho_1)=1$,
and $\mathcal{C}_{l_1}(\rho_2)=\frac{2\sqrt{6}+1}{4}>\mathcal{C}_{l_1}(\rho_1)=1$.
Thus the coherence measures $\mathcal{C}_{tr}^{\prime}$ and $\mathcal{C}_{l_1}$ give a different ordering in this case,
hence also give a different ordering of states in $d$-dimensional quantum systems when $d\geq3$.

Noting that $\mathcal{C}_{l_1}(\rho_m)=(d-1)p$ and $\mathcal{C}_{tr}^{\prime}(\rho_m)=p$ are both increasing functions with respect to $p$,
they provide the same ordering for $\rho_m$.
If we consider the density matrices which have the block-diagonal form under the incoherent basis, and the dimension of each block is at most 2,
then the coherence measures $\mathcal{C}_{l_1}$ and $\mathcal{C}_{tr}^{\prime}$ also give the same ordering.

\medskip

\noindent\textbf{Ordering states with $\mathcal{C}_{tr}^{\prime}$ and $\mathcal{C}_{r}$}\smallskip

Consider the coherence measures $\mathcal{C}_{tr}^{\prime}$ and $\mathcal{C}_r$ for arbitrary $d$-dimensional quantum systems.
When $d=2$, $\mathcal{C}_{tr}^{\prime}(\rho)=\mathcal{C}_{l_1}(\rho)$, it has been proved in \cite{Liu} that
$\mathcal{C}_{tr}^{\prime}$ and $\mathcal{C}_r$ give rise to a different ordering.
When $d\geq3$, similar to the discussion in Sect. 3.2, one can demonstrate that
$\mathcal{C}_{tr}^{\prime}$ and $\mathcal{C}_r$ also give rise to a different ordering.

There also have been sets of quantum states such that $\mathcal{C}_{tr}^{\prime}$ and $\mathcal{C}_r$ provide the same ordering.
Note that the coherence measures $\mathcal{C}_{l_1}$ and $\mathcal{C}_{r}$ have the same ordering for all single-qubit states with a fixed mixedness.
Thus $\mathcal{C}_{tr}^{\prime}$ and $\mathcal{C}_r$ also provide the same ordering in this case, since $\mathcal{C}_{tr}^{\prime}(\rho)=\mathcal{C}_{l_1}(\rho)$ for all single-qubit states.
Besides, similarly to the previous discussion, $\mathcal{C}_{tr}^{\prime}$ and $\mathcal{C}_r$ give the same ordering for $\rho_m$.

We now extend our discussion to other coherence measures -- the coherence of formation and the Tsallis relative $\alpha$-entropies of coherence.
The coherence of formation is defined as
$\mathcal{C}_f(\rho)=\underset{\{p_i,\varphi_i\}}{\min}\Sigma_ip_i\mathcal{S}(|\varphi_i\rangle\langle\varphi_i|_{diag})$,
where $\rho=\Sigma_ip_i|\varphi_i\rangle\langle\varphi_i|$ is any pure state decomposition of $\rho$.
The Tsallis relative $\alpha$-entropies of coherence is defined as
$\mathcal{C}_\alpha(\rho)=\underset{\ \delta\in\mathcal{I}}{\min}\mathcal{D}_\alpha(\rho\|\delta)=
\frac{r^\alpha-1}{\alpha-1}$, where $r=\Sigma_i\langle i|\rho^\alpha|i\rangle^{\frac{1}{\alpha}}$, $\alpha\in(0,1)\cup(1,2]$.

It has been shown that for single-qubit states, $\mathcal{C}_{l_1}$, $\mathcal{C}_{f}$ give the same ordering \cite{Liu}.
Thus the four measures $\mathcal{C}_{l_1}$, $\mathcal{C}_{tr}^{\prime}$, $\mathcal{C}_{g}$ and $\mathcal{C}_{f}$
provide the same ordering of single-qubit states.
Moreover, we claim that $\mathcal{C}_\alpha$ and $\mathcal{C}_g$, $\mathcal{C}_\alpha$ and $\mathcal{C}_{tr}^{\prime}$, as well as $\mathcal{C}_\alpha$ and $\mathcal{C}_f$ do not generate the same ordering of single-qubit states when $\alpha=\frac{1}{2}$ and $2$,
since in this case, $\mathcal{C}_\alpha$ and $\mathcal{C}_{l_1}$ generate a different ordering \cite{Tsa}.
The fact that $\mathcal{C}_\alpha$ and $\mathcal{C}_r$ give rise to a different ordering of single-qubit states has also been proposed in Ref. \cite{Tsa}.

In Ref. \cite{mix}, the authors studied the coherent-induced state ordering with fixed mixedness.
They proved that $\mathcal{C}_{l_1}$, $\mathcal{C}_{r}$ and $\mathcal{C}_{\alpha}$ give the same ordering of single-qubit states with a fixed mixedness.
Thus, we get that with a fixed mixedness, the coherence neasures $\mathcal{C}_{l_1}$, $\mathcal{C}_{r}$, $\mathcal{C}_{g}$, $\mathcal{C}_{tr}^{\prime}$, $\mathcal{C}_{f}$ and $\mathcal{C}_{\alpha}$ give the same ordering of single-qubit states.
Therefore the problem of ordering single-qubit states with these six coherence measures is completely solved.

\noindent\textbf{Conclusion}
We have investigated the issue of ordering states with the $l_1$ norm of coherence, the relative entropy of coherence, the geometric
measure of coherence and the modified trace distance measure of coherence.
For single-qubit states, the $l_1$ norm of coherence, the modified trace distance measure of coherence and the geometric coherence give the same ordering.
We also have shown that the relative entropy of coherence and the geometric measure of coherence do not give the same ordering of single-qubit states.
Furthermore, for high-dimensional quantum systems, each pair of the four measures $\mathcal{C}_{l_1}$, $\mathcal{C}_{r}$, $\mathcal{C}_{g}$ and $\mathcal{C}_{tr}^{\prime}$ give a different ordering.
However, for some special classes of quantum states, each pair of these measures may provide the same ordering.
For instance, we have shown that they give the same ordering for a class of maximally coherent mixed states $\rho_{m}$.
We also have completely solved the problem of ordering single-qubit states with the above four measures, coherence of formation and the Tsallis relative $\alpha$-entropies of coherence.
For each pair of the four measures $\mathcal{C}_{l_1}$, $\mathcal{C}_{r}$, $\mathcal{C}_{g}$ and $\mathcal{C}_{tr}^{\prime}$, we also give some sets under
which they give the same ordering.
It should be noted that, as it was shown in \cite{Maziero}, the "non-equivalence" between the relative entropy and $l_1$-norm coherence is due to the dependence of the first on the density matrix populations, in contrast to the last.
However, we cannot follow the idea in this paper since the two coherence measures $C_g$ and $C_{tr}^{\prime}$ have no analytical expressions in general.
In other words, we do not know whether or not these two coherence measures are dependence of density matrix populations.
Further efforts can be made towards whether or not there exist other coherence measures which generate the same ordering of qudit states.

\vspace{2.5ex}
\noindent{\bf Acknowledgments}\, \,
This work is supported by the NSFC under number 11675113.

\end{document}